\documentclass[eqsecnum,aps,prd,nofootinbib,twocolumn,a4paper]{revtex4}
\usepackage{times} 
\usepackage{enumerate}
\begin{document}

 \def\beq{\begin{equation}}
 \def\endeq{\end{equation}}
 \def\bea{\begin{eqnarray}}
 \def\endea{\end{eqnarray}}
 \def\journaldata#1#2#3#4{{\it #1 } {\bf #2:} #3 (#4)}
 \def\eprint#1{$\langle$#1\hbox{$\rangle$}}
 \def\lto{\mathop        {\hbox{${\lower3.8pt\hbox{$<$}}\atop{\raise0.2pt\hbox{$\sim$}}$}}}

\title{Resource Letter BH-2: Black Holes}

\author{Elena Gallo}\email{egallo@mit.edu}\affiliation{Massachusetts Institute of Technology, Kavli Institute for Astrophysics and Space Research, 70 Vassar St., Bldg 37-685, Camdridge, MA 02139}\altaffiliation{Hubble Fellow}\author{Donald Marolf}\email{marolf@physics.ucsb.edu}
\affiliation
{Department of Physics, University of California Santa Barbara, CA 93106-9530}

\begin{abstract}
This Resource Letter is designed to guide students, educators, and researchers
through (some of) the literature on black holes.  We discuss both the physics and
astrophysics of black holes.  We emphasize breadth over
depth, and review articles over primary sources.  We include resources ranging
from nontechnical discussions appropriate for broad audiences to technical
reviews of current research.  Topics addressed include classification of
stationary solutions, perturbations and stability of black holes, numerical
simulations, collisions, the production of gravity waves, black-hole
thermodynamics and Hawking radiation, quantum treatments of black holes, black
holes in both higher and lower dimensions, and connections to nuclear and
condensed-matter physics. On the astronomical end, we also cover the physics of gas
accretion onto black holes, relativistic jets, gravitationally red-shifted
emission lines, evidence for stellar-mass black holes in binary systems and
supermassive black holes at the centers of galaxies, the quest for intermediate-mass black holes, the assembly and merging history of supermassive black
holes through cosmic time, and their affects on the evolution of galaxies.

\end{abstract}

\maketitle

\section{Introduction}

The status of black-hole physics has changed dramatically since the
publication of the first black-hole Resource Letter, ref. below.  The
observational evidence for black holes is now plentiful and strong.  Moreover,
black holes have progressed from a curiosity of mathematical physics to a {\it
tool} for building or studying models in other branches of physics.  Examples
range from feedback mechanisms that may control the formation and evolution of
galaxies (see refs. \ref{dimatteo}, \ref{feedback}) to string-theoretic connections to
nuclear and condensed matter physics (see refs. \ref{Nature}, \ref{RHIC},
\ref{CM}).  Nevertheless, black holes remain fascinating objects in their own
right that may yet divulge deep clues to further revolutions in fundamental
physics.

We have endeavored to provide a guide to this extraordinarily diverse literature, and to include resources at a variety of levels.  Some advanced resources will require additional specialized knowledge of general relativity, astrophysics, or quantum-field theory.  We hope that bringing together many levels of material will allow students of black-hole physics to find the right entry point for their background and teachers of black-hole physics to extract useful material for their courses.

Given the breadth of our subject, it is impossible to be complete.  We have by no means attempted to do so.  Experts will find many of their favorite resources, and perhaps even their favorite topics to be missing entirely.  We have concentrated on a ``few'' review articles and  seminal papers, though we have tried to include sufficient commentary to help the user locate the most useful resource.  Where possible, we have chosen review articles over original works and modern treatments over classic presentations.   We made this decision for purely practical reasons, and we encourage the reader to consult Steve Detweiler's original black-hole resource letter (ref. below) for an excellent guide to the historical literature.  Indeed, we view our work as a supplement to his, not as a replacement.

\begin{enumerate}[{\bf 1{.}}]

\item { ``Resource Letter BH-1: Black Holes,''} S. Detweiler,
Am. J. Phys. {\bf 49} 394-400 (1981).

\end{enumerate}

Resources in the following sections are organized first by level: popular material appears in section \ref{ele} while most  U.S. undergraduate material appears in section \ref{ug}.  Graduate general-relativity textbooks are found in section \ref{gen} while more specialized material is presented in sections \ref{ABH}-\ref{last}.  However, some of the material in these latter sections is also of intermediate level,  accessible to advanced undergraduate students.
We have then attempted to sort the more specialized papers by topic.  This task is far from straightforward, as many excellent reviews overlap several obvious categories.   The result is that many resources are cross-referenced, typically by adding a note to ``see also ref. nnn."  The reader is well-advised to do so.

We have arranged the specialized material as follows:  Section \ref{ABH}
contains resources dealing with black holes in astrophysics.  This includes
observations of black holes, models of astrophysical processes involving black
holes, and the implications of these models.


Section \ref{SDT} then addresses well-established mathematical physics describing the statics, dynamics, and thermodynamics of black holes.  The focus here is on classical black-hole physics such as the treatment and classification of black holes in general relativity and related theories (in a variety of dimensions), their perturbations and stability properties, their formation and interaction through collisions, the production of gravitational radiation and the relation to gravity-wave detectors, and black-hole thermodynamics.    In addition, however, section \ref{thermo} includes broadly accepted material on quantum field theory in curved spacetime and the associated Hawking radiation.  Such subjects are deeply connected to the thermodynamics of  black holes and are addressed in much of the same literature.  Discussions of the possible production of black holes in particle collisions has also been included in section \ref{dnot4}.

In contrast, those references that concern the deep quantum microphysics of black holes have been placed in section \ref{micro}.  Such resources are more speculative, at least in the sense that they typically  make assumptions about the quantum nature of gravity that are beyond the reach of current experiments.  We therefore separate this material from that of section \ref{thermo}, even though Hawking radiation is again a common theme.  Finally, we have collected connections to nuclear and condensed-matter physics in the short section \ref{last}.   Such connections include the use of established fluid dynamics and condensed-matter physics to describe certain analogues of black holes,   as well as current attempts to use black-hole physics (and string theory) to describe as yet poorly understood effects in both nuclear and condensed-matter physics. Here the emphasis is again on classical black-hole physics but in novel, and sometimes somewhat speculative contexts.

\subsection{Basic Resources and Notation}

Perhaps the most important information we can convey is how the reader might
find and access useful works that will appear only {\it after} the publication of
this Resource Letter.  Web-based search engines continue to make this more and
more possible, but it is worthwhile to point out a few places to start.  When
entering a new subject, it is natural to begin with a review article.  Several
journals are dedicated entirely to such reviews, and one may begin by
searching their websites.  Certain online tools are also of particular help
in locating reviews.  Some suggestions are:

\begin{enumerate} [{\it 1{.}}]

\item {\it Living Reviews in Relativity}.
\newline This online journal publishes only review articles on gravitational physics.  As you might guess, a large fraction of its contents deal with black holes, making it a prime resource.   Indeed, many of the articles described below are from this journal.

\item{\it Reviews of Modern Physics}.
\newline A review journal published by the American Physical Society.  Topics covered span all of physics.

\item {\it Physics Reports.}
\newline A review journal by Elsevier.  Topics covered span all of physics.

\item {\it HEP Reviews at SPIRES},
{\url{www.slac.stanford.edu/spires/reviews/}}.
\newline An ongoing project to
index online reviews in high-energy and theoretical physics.  Black holes are
currently category Id2.  This website also explains how to perform a more
general search of the SPIRES database for relevant review articles.

\item {\it The SAO/NASA Astrophysics Data System Abstract Service (ADS)},
{\url{adsabs.harvard.edu/abstract_service.html/}}.
\newline Hosted by the Smithsonian Astrophysical Observatory for NASA, this
website is the most powerful browser for astrophysics-related material.  It
allows to search for, {\it e.g.}: specific authors (or list thereof), refereed
vs. non-refereeed publications, works mentioning your favorite astronomical object, as
well as title keywords, publication year, number of citations, etc. It covers
a wide spectrum of levels, from scholarly research papers, to Ph.D. thesis, to
conference proceedings. A full description of the search criteria is given at:
{\url{doc.adsabs.harvard.edu/abs_doc/help_pages/basicsearch.html#Basic_Query_Form}}.
ADS has become an irreplaceable instrument in the workday of every
astrophysicist.

\item {\it Annual Reviews of Astronomy and Astrophysics},
{\url{arjournals.annualreviews.org/loi/astro/}}.
\newline ARAA publishes comprehensive
reviews written by well-known authors.  They are an excellent starting point
for the interested reader who wishes to approach a new astrophysical subject
at the intermediate-advanced level.

\end{enumerate}

The resources that we list below should not be hard to locate.
Most appear in journals that can be found in a typical university library or online.  In addition, most also include an ``Arxiv identifier" either of the form
 [arXiv:astro-ph/yymmnnn] or [arXiv:yymm.nnnn [astro-ph]], with astro-ph
perhaps replaced by qr-qc, hep-th, or hep-ph.  These reference numbers refer
to the online physics archive currently maintained by the Cornell University
Library: {\url{arxiv.org/}}.  Many papers can be found here by using the search feature.  To find the precise desired work, simply search by ``identifier" and enter it into the search box:  {\it e.g.},  astro-ph/yymmnnn in the first (older style) numbering system or simply yymm.nnnn in the second system.  Here yy denotes the year, mm the month, and nnn or nnnn the number assigned to the work when it was uploaded to arxiv.org.

As a final remark, the reader should be aware that two acronyms are frequently used in the text below:  General Relativity is abbreviated ``GR," and ``AGN" refers to Active Galactic Nuclei. We have otherwise attempted to make the comments following each resource as approachable as possible by the general reader.

\section{Popular Material}
\label{ele}

Material in this section should be accessible to persons with a high school or college-level education without specialized training in physics.   The goal of these works is typically to convey the general idea of the subject, without going into technical details.  Certain resources are nevertheless somewhat demanding of the reader.  

\subsection{Popular Books}

\begin{enumerate}[{\bf 1{.}}]
\addtocounter{enumi}{1}

\item {\bf Black Holes and Time Warps: Einstein's Outrageous Legacy},
K. S. Thorne (W.W. Norton, New York, 1994).  Although it is now 14
years old, this book remains the most popular
one on this subject.  First and foremost, it is an accessible
introduction to basic issues:  What is a black hole?  How are they
made?  What happens when you fall in? The text begins by explaining the essential ideas in relativity and proceeds through the
development of black-hole physics itself.  The approach is
historical, and the author's personal involvement in this history
cements a gripping story line. But it is the style of the text that
captivates many readers.  The author's charming personality shines
through.  The final chapter on wormholes should be regarded as more
speculative material, but the rest of the book focuses on
well-established physics. (E)

\item {\bf Black Hole Physics: Basic Concepts and New Developments},
V. P. Frolov and I. D. Novikov (Springer, New York, 1998). An encyclopedic,
self-contained text on the physics of black holes, with a special focus on
black holes in astrophysics (I).

\item {\bf Gravity's Fatal Attraction: Black Holes in the Universe},
\label{BR}
M. Begelman and M. Rees (Scientific American Library, New York, 1995).  Do ``spacetime
singularities'' actually exist in nature? Have we found them? Begelman and
Rees tell us the
captivating story of the scientific endeavors --and related anecdotes--
that led to the discovery of black holes. Powerful, inspiring, and fun to read. (E)

\item {\bf General Relativity from A to B}, R. Geroch (U. Chicago Press, Chicago, 1978).
Developed as lectures to nonscience majors at the University of Chicago, Geroch's book gives a solid conceptual explanation of special and GR.  Some math is used.  This book requires more effort than those above, but delivers a deeper understanding of relativity. (E/I)

\item {\bf Was Einstein Right? Putting General Relativity to the Test},
C. M. Will (BasicBooks, New York, 1993).  Most popular GR books emphasize the theory and conceptual underpinnings of the theory.  In contrast, Will's book focusses on the many empirical tests of this theory, including Hulse and Taylor's Nobel-prize winning verification of the decay of neutron star orbits via gravitational radiation.  (E/I)

\item {\bf Relativity}, A. Einstein,  (Three Rivers Press, New York, 1961).   Most later expositions of GR are based at least in part on this book, Einstein's attempt to describe GR to anyone with a standard education.  In his own words, ``despite the shortness of the book, a fair amount of patience and force of will" is required of the reader.  Nevertheless, the book is an invaluable opportunity to learn at least a bit of relativity from Einstein himself. (I)

\end{enumerate}

\subsection{Websites}

We highly recommend the websites below, all at the elementary level.  Of course, websites are in constant flux.  Check your favorite search engine.

\begin{enumerate}[{\bf 1{.}}]

\addtocounter{enumi}{7}

\item \label{wiki}
{\url{en.wikipedia.org/wiki/Black_hole/}} \newline
{\bf Black Hole}, Wikipedia.  As usual, this Wikipedia entry is surprisingly broad.  The reference list is excellent. (E)

\item \label{hubblesite}
{\url{hubblesite.org/explore_astronomy/black_holes/}} \newline
{\bf Black Holes: Gravity's Relentless Pull},  Hubble Site.  Packed with
graphics, simulated ``experiments,"  and animations, this professional site
from NASA and the Space Telescope Science Institute provides an exciting tour
of black-hole physics. (E)

\item \label{Jillian} {\url{www.gothosenterprises.com/black_holes/}}
\newline {\bf Jillian's Guide to Black Holes}, J. Bornak.
A charming, informal, and plain-speaking introduction to black holes and GR.  Written by a black-hole fan as an undergraduate project.  Nonprofessional in a delightful way and a marvelous introduction to the basic physics. (E)

\item \label{space} {\url{www.space.com/blackholes/}}
\newline {\bf All About Black Holes}.
Lovely videos and simulations, as well as black-hole news items. (E)

\item \label{RN}
{\url{antwrp.gsfc.nasa.gov/htmltest/rjn_bht.html/}}\newline
{\bf Virtual Trips to Black Holes and Neutron Stars}, R.
Nemiroff.  The way black holes curve spacetime makes light do crazy things.  What would you see if you approached a black hole?  Check this classic website to find out. (E)

\item \label{Hamilton} {\url{casa.colorado.edu/~ajsh/schw.shtml/}}\newline
{\bf Falling Into a Black Hole}, A. Hamilton.  More modern simulations of trips to and inside black holes, but with a somewhat confusing portrayal of the horizon. (E)

\item \label {stellar}
{\url{www-xray.ast.cam.ac.uk/xray_introduction/Blackholebinary.html/}}
\newline {\bf Black Holes and X-ray Binaries}. Black holes and astrophysics from the Institute of Astronomy, Cambridge. (E)

\item  library.thinkquest.org/25715/index.htm/
\newline {\bf Event Horizon}.  Black-hole basics with great simulations. (E)

\item \label{xrays}
{\url{www.astro.umd.edu/~chris/Research/X-rays_and_Black_holes/x-rays_and_black_holes.html/}}
\newline {\bf X-rays and Black Holes}, C. Reynolds, University of Maryland.
A brief, informal and yet accurate description of the status of research on
astrophysical black holes, explained for the nonexperts by a forefront leader in the
field. (E)

\item
{\url{www.youtube.com/watch?v=hoLvOvGW3Tk}}
\newline {\bf Black Holes, Neutron Stars, White Dwarfs, Space and Time}, YouTube.
Simulations of black holes feeding and being born through collisions, all set to an exciting sound track. (E)

\item \label {NoEsc} {\url{amazing-space.stsci.edu/resources/explorations/blackholes/}}
\newline
{\bf Amazing Space}, Space Telescope Science Institute.
This website provides black-hole resources for teachers and students. (E)

\end{enumerate}

\subsection{Popular Articles}

The reader can find a number of interesting black-hole articles in
popular-science publications such as  {\it Scientific American},
{\it Physics World}, {\it Physics Today}, {\it New Scientist} and,
at a slightly more advanced level,  {\it Science} and {\it Nature}.
Some fairly recent articles are listed below, but there is no doubt that more will
appear in the near future.    We have sorted these articles roughly into those that emphasize the astrophysics of black holes and those emphasizing fundamental aspects of theoretical physics.

\medskip

\medskip

\medskip

{\bf ~Astrophysics: }

\begin{enumerate}[{\bf 1{.}}]

\addtocounter{enumi}{18}

\item \label{BHB}  { ``Black Hole Blowback,''} W. Tucker, H. Tananbaum, A. Fabian,
{Scientific American}, {\bf 296}, 42-49 (Feb. 2007).   Black holes and active galactic nuclei as possible
drivers of feedback in galaxy cluster evolution. (E)

\item {``Unmasking Black Holes,''} J. Lasota, Scientific American, {\bf 280}, 30-37 (May
1999).  On how we might directly detect black-hole horizons. (E)

\item
{``The Galactic Odd Couple,''} K. Weaver, Scientific American, {\bf 289}, 34-41 (July 2003).
Is there a connection between black holes and bursts of start formation? (E)

\item {``The Midlife Crisis of the Cosmos,''}  A. J. Barger, Scientific
American, {\bf 292} 32-39 (Jan. 2005).
The authors describe the present and past of star and black-hole formation in our universe. (E)

\item {``The Brightest Explosions in the Universe,''}  N. Gehrels, L. Piro
and P. J. T. Leonard, Scientific American, {\bf 287} 84-91(Dec. 2002). Connecting the birth of black holes to enormous explosions called gamma-ray bursts. (E)\\

\medskip

\noindent
{\bf Fundamental Physics:}

\item \label{QBH} {``Quantum Black Holes,''} B. J. Carr and S. B. Giddings, Scientific American, {\bf 292}, 48-55 (May 2006).  Might we soon be able to make black holes in the laboratory? (E)

\item {``The Reluctant Father of Black Holes,''} J. Bernstein, Scientific
American, {\bf 274}, 80-85 (June 1996).
The early history of our understanding of black holes
and of Einstein's resistance to accepting them.  (E)

\item  {``An Echo of Black Holes,''} T. A. Jacobson and R. Parentani,
Scientific American, {\bf 293} 68-75 (May 2006).
 Flowing fluids in the laboratory can act much like black holes, and even have a form of Hawking radiation.  (E)

\item {``Black Hole Computers,''}  S. Lloyd and Y. J. Ng, Scientific
American, {\bf 291}, 52-61 (Nov. 2004).
Black Holes and Information Theory. (E)

\item {
``Information in the Holographic Universe,''} J. D. Bekenstein,
Scientific American, {\bf 289}, 58-65 (Aug 2003). How black holes may be related to
fundamental bounds on the information that objects can hold. (E)

\item \label{IoG} {``The Illusion of Gravity,''}  J. Maldacena, Scientific
American, {\bf 293}, 57-63 (Nov. 2005).
String theory has holographic properties that seem deeply linked to black hole physics.  (E)

\item {``Reality-Bending Black Holes,''} Scientific American,
Special Edition, March 2007.  This volume collects references 
\ref{BHB}-\ref{IoG} above, along with other articles on related topics. (E)

\end{enumerate}

\section{Undergraduate Material}

This section describes resources at the advanced (U.S.)
undergraduate level.  Some entries should be accessible to any
student who has completed a one-year calculus-based physics course,
while others require background in
intermediate classical mechanics, quantum mechanics, and
electromagnetism.  The articles in section \ref{ugart} address special topics not covered by standard textbooks, but which are nevertheless approachable by readers without detailed knowledge of GR.   Students who wish to establish a working
understanding of black holes should consult one of the undergraduate
textbooks in section \ref{ugbook}.

 \label{ug}

\subsection{Articles}
\label{ugart}

\begin{enumerate}[{\bf 1{.}}]\addtocounter{enumi}{30}

\item {``Black holes: the inside story,''}
S. Droz, W. Israel and S. M. Morsink, Physics World, {\bf 9}, 34-37 (1996). One
of a very few articles with a good description of black-hole interiors.  A key
point is the instability of the so-called inner horizon of charged and
rotating black holes.  This instability forbids the use of black holes as
passageways to other universes.  The article is also posted online at
{\url{www.phys.ualberta.ca/~morsink/blackhole/}} in both English and French
translations. (E)

\item \label{BHD1}  { ``Spacetime embedding diagrams for black holes,''}
D. Marolf,
  Gen.\ Rel.\ Grav.\  {\bf 31}, 919-944 (1999).
  [arXiv:gr-qc/9806123].  The curved space{\it time}
of a Schwarzschild black hole can be visualized by drawing the so-called $rt$-plane as
a curved surface inside flat 2+1 Minkowski space.  The main text
is accessible to those with some training in special
relativity, though some knowledge of GR is required
to follow the construction (in the appendix) of the diagrams from
the black hole metric. (I)

\item
  { ``Gravity and the quantum,''} A.~Ashtekar,
  New J.\ Phys.\  {\bf 7}, 198-198 (2005)
  [arXiv:gr-qc/0410054].
  Ashtekar gives a broad overview of loop quantum gravity for
  the nonexpert.  After introducing the basic ideas, section IIIB
  addresses black hole entropy  in loop quantum gravity.  The
  article should be accessible to students familiar with
  undergraduate quantum mechanics, the description of
  electromagnetism via 4-vector potentials, and some knowledge of
  GR. (I)

\item \label{EW}  { ``Black Holes and Quark Confinement,''}
E. Witten, Current Science {\bf 81}, 1576-1581 (2001). Witten
takes the reader on a tour of ideas ranging from
strings to quarks to black holes and the AdS/CFT
duality.  The climax is the fusion of these ideas into a stringy
derivation of quark confinement in theories mathematically similar
to, but still different than the QCD that describes the strong
interactions of our universe. (I)

\item \label{SW} { ``A Microscopic Theory of Black Holes in String
Theory,''}
S. R. Wadia, Current Science {\bf 81} (12), 1591-1597 (2001). After
providing an undergraduate-level description of black-hole
thermodynamics and Hawking radiation, Wadia describes the extent to
which these phenomena are currently described by string theory.  The
concluding section includes a useful set of open questions.  (I)

\item \label{Nature} { ``Theoretical physics: A black hole full of answers,''} J. Zaanen, Nat. {\bf 448}, 1000-1001 ( 2007).
Zaanen briefly describes current attempts to use black holes and string theory as mathematical tools to explain the physics of both high-temperature superconductors and quark-gluon plasmas.  (I)

\end{enumerate}

\subsection{General Relativity Textbooks}

A number of excellent GR textbooks are now available at the  U.S. undergraduate level, each with a good treatment of black-hole physics.  The texts differ in terms of both perspective and level of
sophistication, but each one provides what is in principle a
self-contained introduction.  We have ordered the books
below roughly in terms of level of presentation.

\label{ugbook}
\begin{enumerate}[{\bf 1{.}}]\addtocounter{enumi}{36}

\item \label{GroundUp} {\bf Gravity from  the ground up,} B. Schutz (Cambridge University Press, Cambridge, 2003).
This excellent book emphasizes {\it gravity}, and its interactions with astrophysics.  Assuming only high-school math and physics, Schutz addresses the solar system, the birth and life cycle of stars, relativity, cosmology, gravity waves, and gravitational lenses.  The chapter on black holes summarizes a wide range of physics from accretion and black hole growth to Hawking radiation and black-hole entropy. (E/I)

\item \label{Notes} {\bf Notes on Relativity and Cosmology}, \newline  D.
Marolf (2003).  Available on-line at
{\url{www.physics.ucsb.edu/Smarolf/MasterNotes.pdf}}.  Designed to convey understanding of relativity using only elementary calculus,
these notes first thoroughly treat accelerated reference frames in special
relativity.  Spacetime diagrams are favored over calculations.  GR is
introduced by piecing together such frames using the equivalence
principle.  The chapter on black holes emphasizes similarities between horizons and accelerated frames in flat spacetime, and also summarizes ref. \ref{BHD1} above.   (E/I)

\item \label{Taylor} {\bf Exploring Black Holes: Introduction to General Relativity},
E. F. Taylor and J. A. Wheeler (Addison Wesley Longman, San Francisco, 2000).
Taylor and Wheeler focus on a nuts-and-bolts exploration of black-hole
solutions as an introduction to GR.  The text requires algebra and basic
calculus but does not assume prior knowledge of physics or relativity.  The
emphasis is on the experience of an observer near a black hole.  (I)

\item \label{EllisWilliams} {\bf
Flat and Curved Space-times, 2nd Edition},  by G. F. R. Ellis and R.
Williams (New York, Oxford University Press, 2000). This excellent
book on special and general relativity includes a survey of
basic black-hole results including no-hair theorems, energy
extraction, and thermodynamics. Ellis and Williams assume knowledge
of basic calculus and introduce other mathematics as needed.  (I)

\item \label{hartle} {\bf Gravity, An Introduction to Einstein's General Relativity},
J. B. Hartle  (Addison Wesley, San Francisco, 2003).  We recommend
Hartle's advanced undergraduate text to beginners and experts alike.
 It provides a thorough introduction to GR with a
 delightful emphasis on both physical concepts and experimental tests.
 Hartle treats black holes in some depth, discussing orbits, energy extraction, and the detailed solutions.  (I)

\end{enumerate}

\section{Advanced Material: General}

\label{gen}

Good introductions to black-hole physics can be found in many
general-relativity textbooks, of which a large number are available at the U.S.
graduate-student level.  Some of our favorites are described below.
Unless otherwise noted, these books require a thorough understanding
of undergraduate-level classical mechanics but do not presume
knowledge of more advanced material.    Sections  \ref{ABH}-\ref{last} provide a guide to particular topics in black-hole physics and include a few more specialized texts.

\begin{enumerate}[{\bf 1{.}}]\addtocounter{enumi}{41}

\item \label{SchutzGreen} {\bf A First Course in General Relativity}, B. F. Schutz (Cambridge University Press, Cambridge, 1985).
A text at the advanced undergraduate or early graduate-level with an excellent treatment of stress-energy tensors and a brief introduction to black-hole physics.
 (I/A)

\item \label{Carroll}  {\bf Spacetime and Geometry: An Introduction to General
Relativity,} S. Carroll (Benjamin Cummings, San Francisco, 2003).  A modern beginning graduate-level text that emphasizes links to field theory.  Carroll's two chapters on black-hole physics provide a readable introduction to the subject. (A)

\item \label{WaldBook} {\bf General Relativity}, R. M. Wald (U. Chicago Press, Chicago, 1984).  Wald's book is the standard text for a mathematical-physics-based advanced graduate course on general relativity.  (A)

\item \label{MTW} {\bf Gravitation}, C. Misner, K. S. Thorne, J. A. Wheeler (W.H. Freeman and Company, New York, 1973).  Misner, Thorne, and Wheeler provide an encyclopedic reference for GR as of the early 1970s.  (A)

\item \label{Weinberg} {\bf Gravitation and Cosmology: Principles and Applications of the General Theory of Relativity}, S. Weinberg (John Wiley, New York, 1972).  A good text for those who wish to do calculations.  Weinberg's book emphasizes equations over geometric intuition and provides a solid connection to cosmology. (A)

\item \label{HE} {\bf The large scale structure of space-time}, S. W. Hawking and G. F. R. Ellis (Cambridge University Press, Cambridge, 1973).  The classic reference on causal structures, singularity theorems, and mathematical physics aspects of general relativity. (A)

\end{enumerate}

\section{Astrophysical Black Holes}
\label{ABH}

It is now thought that almost all galaxies contain ``supermassive" black holes at
their centers, millions or even billions of times more massive than the
Sun. Some of these objects power the most energetic (nonexplosive) phenomena in
the universe, such as active galactic nuclei (AGN) and quasars. Others, like the
black hole at the center of the Milky Way, are much more quiet.
Galaxies are also thought to contain many examples of stellar-mass black
holes, with masses a few times greater than that of the Sun, which are the
evolutionary endpoint of very massive stars.

While we cannot observe black holes directly, we can observe and measure
their effects on the surroundings. In fact, black holes are extremely
efficient at converting the gravitational potential energy of in-falling gas
into radiation. Except under special circumstances, the gas that falls into a
black hole (be it interstellar/intergalactic gas, or supplied by a companion
star) does not plunge in directly; instead it forms what is known as an
accretion disk. The inner disks of supermassive black holes reach thousands of
degrees Kelvin, while stellar-mass black holes can heat their disks to
millions of degrees, where they emit in the X-ray part of the spectrum.

After listing a few basic texts in section \ref{Atexts}, in sections
\ref{stellarMBH}-\ref{IMBH} we organize other resources by whether the relevant
black holes are stellar mass, supermassive, or of intermediate mass scale(between the
former two). Material concerned with specific astrophysical sources, most notably
the supermassive black hole in the center of our own galaxy, appears
separately in section \ref{BHCS}.

\subsection{Astrophysics: Basic Textbooks and Reviews}
\label{Atexts}

See ref. \ref{BR} from section~\ref{ele} A.
\begin{enumerate}[{\bf 1{.}}]
\addtocounter{enumi}{47}

\item{ ``Probes and Tests of Strong-Field Gravity with Observations in the
Electromagnetic Spectrum,''} D. Psaltis, Living Reviews in Relativity (2008),
[arXiv:0806.1531]. This review serves as a nexus between the previous
sections of this Resource Letter and the following ones, which cover the
observable aspects of black holes. Psaltis summarizes the current prospects for
using astrophysical sources hosting black holes as tools for testing the
predictions of GR in the strong-field regime, and does so in a concise and
clear manner. (I)

\item {\bf Accretion Power in Astrophysics}, J. Frank, A. King, and
D. Raine (Cambridge University Press, Cambridge, 2002).
Assuming a basic knowledge of physics, the authors examine the process of
accretion, {\it i.e.} the extraction of gravitational potential energy from matter
falling onto a compact object. A fundamental textbook for astronomy graduate
students. (I/A)

\item \label{super}
{\bf The Galactic Supermassive Black Hole}, F. Melia (Princeton University Press, Princeton, 2007). This book is an excellent starting point for astrophysics
graduate students interested in the black hole associated with galactic center
source Sagittarius A*.  The history of observations is described along with
the observational signatures themselves.  These observations range
from radio to X-ray wavelengths.  Nearby stellar orbits, accretion,
flares, and lensing are discussed.  While primarily written from the
astrophysical point of view, the text devotes one chapter to
explaining elements of GR that are key to understanding black holes
in general. (A)

\item { ``Fluorescent iron lines as a probe of astrophysical black hole
systems,''} C. Reynolds and M. Nowak, Physics Reports, {\bf 377}, 389-466 (2003) [arXiv:astro-ph/0212065].
As gas accretes onto a black hole, its velocity approaches the speed
of light. This should give rise to observable relativistic effects, such as
gravitationally red-shifted emission lines. In this extensive
review, Reynolds and Nowak describe relativistic iron lines as probe of the innermost regions
around accreting black holes. It also features a broad, pedagogical introduction. (I)

\item{  ``Black Holes and Relativistic Jets,''}
R. D. Blandford, Progress of Theor. Phys. Suppl. {\bf 143}, 182-201 (2001) [arXiv:astro-ph/0110394].
Accreting black holes of all mass scales seem to be able to form narrow,
bipolar streams of outflowing matter/energy, as known as jets. Blandford
discusses a variety of viable mechanisms through which mass, angular momentum,
and energy can escape the pull of gravity and be ejected outwards at
relativistic velocities. Particular attention is paid to the connection
between the jet, the accretion disk, and the black hole itself. Written by a
world leader, as well as perhaps the most cited author(ity), in the field.
(I/A)

\end{enumerate}

\subsection{stellar-mass black holes}
\label{stellarMBH}
\begin{enumerate}[{\bf 1{.}}]

\addtocounter{enumi}{52}

\item {\bf Black Holes, White Dwarfs and Neutron Stars: The Physics of
Compact Objects}, S. L. Shapiro and A. Teukolsky (John Wiley \& Sons Inc., New York,
1983).  This excellent textbook brings together different branches of physics
-- from nuclear physics, to hydrodynamics, to special relativity and GR -- to
understand the birth, evolution and death of a massive star under the special
circumstances that lead to the
formation of a binary star system hosting a relativistic compact object. (A)

\item { ``Black holes in binary systems. Observational appearance,''}
N. I. Shakura and R. A. Sunyaev, Astron. Astrophys., {\bf 24}, 337-355 (1973).
In this seminal paper, Shakura and Sunyaev explain how, in a binary system
made of an evolved star that transfers part of its envelope mass to a black hole, the outward
transfer of angular momentum leads to the formation of a disk around the black
hole. (A)

\item { ``Image of a spherical black hole with thin accretion disk,''}
J.-P. Luminet, Astron. Astrophys., {\bf 75}, 228-235 (1979). The first
investigation of the actual optical appearance of a spherical black hole
surrounded by thin accretion disk; the spectral shifts
arising from gravitational and Doppler shifts result in strong asymmetry in
the flux distribution. This calculation can be
generalized to black holes of any mass. (A)

\item { ``Black hole binaries,''}
\label{mccr}
J. E. McClintock and R. A. Remillard, in {\bf ``Compact stellar X-ray sources,''}
edited by W. Lewin and M. van der Klis (Cambridge University Press, Cambridge, pp. 157-214, 2006) [arXiv:astro-ph/0306213].
The reader is introduced to the 40+ known systems hosting candidate stellar-mass black holes in the
Milky Way. About half of them are known to contain an object whose mass exceeds the
maximum theoretical mass for a neutron star, and are referred to as
``dynamically confirmed'' black hole accretors. (I)

\item { ``Relativistic Jets from X-ray binaries,''}
R. P. Fender, in {\bf ``Compact stellar X-ray sources,''} edited by W. Lewin and
M. van der Klis (Cambridge University Press, Cambridge, pp. 381-420, 2006) [arXiv:astro-ph/0303339].
Relativistic jets are among the most spectacular objects in the universe, and
are increasingly being found throughout the Milky Way. Despite being
discovered in the radio band, they have been proven to emit across the entire
electromagnetic spectrum. This review deals with the variety of jets from
galactic stellar-mass black holes (as well as neutron stars), and their
connection with the accretion flow. (I)

\item {
``Advection-dominated Accretion and Black Hole Event Horizons,''} R. Narayan and
J. E. McClintock, to appear in  New Astronomy Reviews [arXiv:astro-ph/0803.0322]. This work summarizes the observational evidence for the presence of an event horizon in stellar-mass black holes by comparing the luminosities of black holes and neutron stars accreting mass at
low rates. If the accretion flow has a low enough density, the gravitational
potential energy can remain trapped in the flow. In the case of a neutron star,
half of that energy will be released as X-rays upon impact on the star
surface. Instead, the energy will be advected across the event horizon in the
case of a black hole. Indeed, this luminosity gap has been observed in X-rays. (A)

\item {
``Gamma-ray bursts from stellar mass accretion disks around black holes,''}
S. Woosley, Astrophys. J. {\bf 405}, 273-277 (1993).
Gamma-ray bursts are the most energetic, explosive astrophysical
events we know of. In a flash lasting few ticks of a clock, more energy is
released than over the entire Sun lifetime.
Woosley's model for the production of (long) gamma-ray bursts involves the formation
of a stellar-mass black hole as the evolutionary endpoint of Wolf-Rayets,
i.e. a particular class of massive stars. (A)
\end{enumerate}

\subsection{supermassive black holes}
\label{stellarSMBH}
\begin{enumerate}[{\bf 1{.}}]

\addtocounter{enumi}{59}

\item {\bf
Active galactic nuclei: from the central black hole to the galactic
environment}, J. Krolik (Princeton University Press, Princeton, 1998).
A comprehensive description of the observational properties of supermassive black
holes that are actively accreting at the center of AGN.
A must-read resource for astronomers, physicists interested
in applications of the theory of gravitation, and graduate students. (E/I)

\item {
``Black Hole Models for Active Galactic Nuclei,''} M. J. Rees,
Annu. Rev. Astron. Astroph., {\bf 22}, 471-506 (1984).  An inspired -- and
inspiring -- work that reviews the body of evidence for supermassive black
holes as the power engine for AGN: quasars, radio galaxies, blazars, and
related objects. (I)

\item { ``Variability of Active Galactic Nuclei,''}
B. M. Peterson, in {\bf Advanced Lectures on the Starburst-AGN Connection}, edited by 
I. Aretxaga, D. Kunth and R. M{\'u}jica (World Scientific,  Singapore, 2001) pp. 3-68.
A comprehensive description of `reverberation mapping' techniques, which allow one to infer the masses of black holes in active galaxies at high redshifts.
This method is based on measuring the time delay between variations in the line and continuum emission from distant AGN. The delay provides an indirect measure of the distance between the supermassive black holes and the regions where the emission lines are being excited. This characteristic scale size constrains the size of the black hole, and thus its mass. (I)

\item {
``Quasars and galaxy formation,''} J. Silk and M. J. Rees, Astron. and
Astrophys. {\bf 331}, L1-L4 (1998) [arXiv:astro-ph/9801013].
A visionary paper on the interplay between the formation of black-hole seeds in
the early universe and the evolution of (proto-)galaxies. (I/A)

\item {
``Supermassive Black Holes in Galactic Nuclei: Past, Present and Future
Research,''} L. Ferrarese and F. Holland, Space Sciences Rev., {\bf 116}, 523-624,
(2005) [arXiv:astro-ph/0411247].  This review discusses the current status of
supermassive black-hole research from an observational perspective. Recent
observations have unveiled empirical scaling relations between the mass of the
central black hole and the global properties of the host galaxy (see
refs. \ref{scaling1},\ref{scaling2}), suggesting that black holes influence
the very formation and evolution of cosmic structures since the dawn of
time. (I)

\item { ``Massive black holes: formation and evolution''}
M. J. Rees and M. Volonteri, in {\bf Black Holes from Stars to Galaxies --
Across the Range of Masses,} edited by V. Karas and G. Matt (Cambridge
University Press, Cambridge, 2007) [arXiv:astro-ph/0701512]. Rees and Volonteri summarize
the progresses and open questions about the formation and growth of
supermassive black holes, as well as the prospects for using them as
laboratories for testing the strong gravity regime. This work includes a list
of the most important (albeit advanced) bibliographic references in the
field. (I/A)

\item \label{dimatteo} { Energy input from quasars regulates the growth and
activity of black holes and their host galaxies}, T. Di Matteo, V. Springel,
L. Hernquist, Nat. {\bf 433}, 604-607 (2005) [arXiv:astro-ph/0502199]. The
evolution of galaxies along with their nuclear supermassive black holes has
been followed by means of detailed numerical simulations.  It is found that a
galaxy-galaxy merger leads to a massive enhancement of the inner accretion
rate, enough to switch on the nuclear black hole as a quasar. In turn, the
energy release by the quasar is able to quench star formation and halt
further growth. This self-regulatory mechanism sets the quasars' duty cycle
and may be able to explain the empirical scaling relations refs. \ref{scaling1}
and \ref{scaling2}. (I)

\item \label{feedback} { ``The many lives of active galactic nuclei: cooling flows, black
holes and the luminosities and colors of galaxies,''} D. J. Croton {\it et al.},
Mon. Not. R. Astron. Soc., {\bf 365}, 11-28 (2006) [arXiv:astro-ph/0508046].
Energy ``feedback'' from supermassive black holes could solve a number of
problems faced by the hierarchical paradigm for the formation and evolution of
cosmic structures at galactic and sub-galactic scales, such as the observed
red colors of massive spheroidal galaxies. Croton et al. couple the output
from state-of-the-art dark-matter simulations to semianalytical modeling to
show that nuclear supermassive black holes may be able to quench the
formation of young blue stars by injecting heat and
mechanical power into the surroundings.  In order for black-hole feedback to
be effective, low levels of prolonged activity are needed, as opposed to
sudden bursts of power. These effects should be observable in the cores of
nearby, formally inactive galaxies. (A)

\item { ``Massive Black Hole Binary Evolution,''} D. Merritt and
M. Milosavljevi\'c, Living Reviews in Relativity (2005),
{\url{http://relativity.livingreviews.org/Articles/lrr-2005-8/}}.
This article reviews the observational evidence (or
lack thereof) for binary supermassive black holes and discusses their
evolution and final fate. Although a compelling case for a bound binary system
is yet to be discovered, a number of sources are believed to host two
supermassive black-hole binaries at projected separations of less than 3000
light-years. (A)

\item {
``LISA observations of rapidly spinning massive black hole binary systems,''}
A. Vecchio, Phys. Rev. D, {\bf 70} 042001, 17pp.  (2004) [arXiv:astro-ph/0304051].
The Laser Interferometer Space Antenna (LISA: {\url{lisa.nasa.gov/}})
is a (planned) space-based gravitational-wave observatory.
Detecting gravitational waves from supermassive black-hole binaries will have
tremendous impact on our understanding of gravity
in the highly nonlinear relativistic regime, as well as the formation and evolution of
galaxies in a cosmological context.
This review focuses on the effects of black-hole spin on the
expected LISA gravitational-wave signal, and describes how accounting for these effects
can substantially reduce the uncertainties on other parameters, such as
the reduced mass. (A)

\end{enumerate}

\subsection{Intermediate-Mass Black Holes}
\label{IMBH}

\begin{enumerate}[{\bf 1{.}}]

\addtocounter{enumi}{69}

\item  { ``Ultraluminous X-Ray Sources in External Galaxies,''}
A. R. King et al., Astrophys. J. {\bf 552}, L109-L112 (2001)
[arXiv:astro-ph/0104333].  With the launch of sensitive X-ray observatories, a
new population of compact, luminous, off-nuclear X-ray sources has been
unveiled in external galaxies. While these so called ``ultraluminous X-ray
sources'' are natural intermediate-mass black-hole candidates, several --less exotic-- explanations are
offered in this paper. (I/A)

\item  { ``A Comparison of Intermediate-Mass Black Hole Candidate
Ultraluminous X-Ray Sources and Stellar-Mass Black Holes,''} J. M. Miller,
A. C. Fabian and M. C. Miller, Astrophys. J., {\bf 614}, L117-L120 (2004)
[arXiv:astro-ph/0406656].
On the same topic as the previous work, this paper focusses instead on the observational
evidence for intermediate mass black holes in a small number of
ultraluminous X-ray sources, namely those where the X-ray continuum spectra appear too cold
to be due to accreting stellar-mass black holes. (I/A)

\item  { ``X-ray observations of ultraluminous X-ray sources,''} T. P. Roberts, Astrophys. and Space Science, {\bf 311}, 203 (2007)
[arXiv:astro-ph/0706.2562].
A brief, critical review on the X-ray properties of ultraluminous X-ray
sources, and their possible physical nature. (I/A)

\end{enumerate}

\subsection{Black Holes: Case Studies}
\label{BHCS}

Throughout this section, we list a number of journal papers that report on
observational milestones on specific black-hole sources, or an ensemble
of them. The two last entries are devoted to the case for a 3-4
million solar-mass black hole at the center of our own Milky Way. This list is
not meant to be complete, but simply to offer a glance at groundbreaking
astrophysical discoveries.

\begin{enumerate}[{\bf 1{.}}]

\addtocounter{enumi}{72}
	
\item { ``Observation of a Correlated X-Ray Transition in Cygnus X-1,''}
H. Tananbaum et al., Astrophys. J. {\bf 177}, L5-L5 (1972).  X-ray astronomy
began only in the early 1960's. The first rocket flight to detect a cosmic X-ray
source was launched in 1962 by a group at American Science and
Engineering. In the 1970s, the dedicated X-ray satellite ``Uhuru'' repeatedly
observed a newly discovered X-ray source in the Cygnus constellation over a
period of 16 months. Its X-ray flux was found to vary significantly, and in
correlation with its radio counterpart. The first ``black-hole state
transition'' (see ref. \ref{mccr}) had just been discovered. (I)

\item {
``X-ray fluorescence from the inner disc in Cygnus X-1,''} A. C. Fabian {\it et al.},
Mon. Not. R. Astron. Soc. {\bf 238}, 729-736 (1989).
It is argued that the broad emission line in the X-ray spectrum of the 10-solar-mass black-hole candidate in Cygnus X-1 is well accounted for by a model
of fluorescent emission from the inner parts of an inclined accretion disk. (I/A)

\item {
``A Superluminal Source in the Galaxy,''} I. F. Mirabel and L. F. Rodriguez, Nat.,
{\bf 371}, 46-48 (1994).  A seminal paper reporting on the discovery of
radio-emitting plasmons traveling with superluminal velocities (apparent velocities greater than the
speed of light) within the Milky Way. The accepted explanation is that the plasmons are ejected in
opposite directions from the central stellar-mass black hole at speeds higher than
0.7$c$.
If the ejection direction is close to the observer's line of
sight, this can lead to apparent superluminal velocities. (E/I)

\item {
``Gravitationally Redshifted Emission Implying an Accretion Disk and Massive
Black-Hole in the Active Galaxy MCG 6-30-15,''} Y. Tanaka {\it et al.}, Nat., {\bf
375}, 659-661 (1995).
Tanaka and collaborators report on the discovery of
relativistic effects in an X-ray emission line from ionized iron in the galaxy
MCG 6-30-15. The line is extremely broad and asymmetric, with most
of the line flux being red-shifted. (I/A)

\item{  ``Formation of the radio jet in M87 at 100 Schwarzschild radii from the
central black hole'' } W. Junor, J. A. Biretta, and M. Livio, Nat., {\bf 401},
891-892 (1999).  This work reports on high-resolution radio observations of the
inner regions of the nearby active galaxy M87, which powers a highly
collimated jet.  The jet radio maps are consistent with the hypothesis that jets
are formed by an accretion disk around the central black hole, which is
threaded by a magnetic field. (I)

\item{  ``Inward Bound---The Search For Supermassive Black Holes In
Galactic Nuclei''}
\label{scaling1}
J. Kormendy and D. Richstone, Annu. Rev. Astron. Astroph. {\bf 33}, 581-581 (1995).
The discovery of an empirical scaling relation between the mass of nuclear
supermassive black holes and the mass of the host galaxy's bulge. (A)

\item{
``A Fundamental Relation between Supermassive Black Holes and Their Host
Galaxies,''}
\label{scaling2}
L. Ferrarese and D. Merritt, Astrophys. J. {\bf 539}, L9-L12,  (2000) [arXiv:astro-ph/0006053]
\& {
``A Relationship between Nuclear Black Hole Mass and Galaxy Velocity
Dispersion,''} K. Gebhardt et al.,  Astrophys. J. {\bf 539}, L13-L16 (2000) [arXiv:astro-ph/0006289].
The discovery (by two different groups) of an empirical scaling relation between the mass of nuclear
supermassive black holes and the velocity dispersion of stars orbiting much
farther than the black-hole sphere of influence. (A)

\item  { ``A 20,000 Solar Mass Black Hole in the Stellar Cluster G1,''}
K. Gebhardt, R. M. Rich and L. C. Ho, Astrophys. J., {\bf 578}, L41-L46 (2002)
[arXiv:astro-ph/0209313]. At the center of a globular cluster, this object represents the
strongest intermediate-mass black-hole candidate to date. Unlike in the
examples listed in section \ref{IMBH}, its presence is inferred from dynamics
rather than from electromagnetic signals. (A)

\item\label{sgr1}{
``The nucleus of our Galaxy,''} R. Genzel, D. Hollenbach and C. H. Townes, Reports
on Progress in Physics, {\bf 57}, 417-479 (1994). The subject of this review is
the central 100 parsecs of our galaxy, with a strong focus on the central few
parsecs. The authors discuss the stellar and interstellar components, the
importance of magnetic and gravitational forces, the evidence for stellar
formation and a central massive black hole, and the origin and nature of
ionization, outflows and interstellar-gas dynamics. (I/A)

\item \label{sgr2} {
``High Proper-Motion Stars in the Vicinity of Sagittarius A*: Evidence for a
Supermassive Black Hole at the Center of Our Galaxy,''} A. Ghez, B. L. Klein, M. Morris, E. E. Becklin, Astrophys. J., {\bf 509}, 678-686 (1998) [arXiv:astro-ph/9807210].
The description of a groundbreaking astrophysics experiment:
The motion of 90
faint stars in the vicinity of the galactic center has been followed over a
period of two
years with the Keck 10m telescope.
Their motion allows for the determination of the enclosed mass through
Kepler's laws. A central dark
mass of about 3 million solar masses is confined to an area about 10 times
smaller than Earth's orbit around the Sun.
The inferred density leads us to the conclusion
that our galaxy harbors a massive central black hole. (A)

\item
{`` Evidence for a Black Hole from High Rotation Velocities in a Sub-Parsec Region of NGC4258 
,''} 	
M. Miyoshi, J. Moran, J. Herrnstein, L. Greenhill, N. Nakai, P. Diamond, and M. Inoue, Nat., {\bf 373}, 127-129 (1995).
Along with the nucleus of the Milky Way (refs. \ref{sgr1},\ref{sgr2}), this galaxy presents the best known case for a super-massive black hole. 
Here the central dark mass is weighed by reconstructing the line-of-sight velocity profile of water-maser-emitting gas orbiting around it. 
Since water-maser emission is a narrow feature, the line-of-sight velocity can be determined from the Doppler shift variations of the observed  maser frequency. The emission regions trace Keplerian orbits which allow one to estimate the enclosed mass as a function of radius, and hence the dark mass density. NGC4258 is thought to host a 40 million solar mass black hole. (I)
\end{enumerate}

\subsection{Image Galleries: Black Holes from Space}

This is all well and fine, one may say. But what is it that we actually
{\it see} when we go up there, outside of Earth's atmosphere?
Browsing the following websites, the reader will be taken through a
journey of the universe as seen by NASA's three ``Great Observatories.''

\begin{enumerate}[{\bf 1{.}}]

\addtocounter{enumi}{83}
\item
\url{hubblesite.org/gallery/album/exotic_collection/black_hole/},\newline {\bf
Space Telescope Science Institute/HST Gallery: Black Holes}. An
amazing gallery of AGN pictures, taken by one the most
powerful optical eyes ever built. NASA's crown jewel, the Hubble Space
Telescope, has provided us with a window to the unexplored universe since 1990. (E)

\item
{\url{chandra.harvard.edu/photo/category/blackholes.html}},\newline {\bf Chandra
X-ray Observatory Photo Album: Black Holes}. Black holes shine in the X-ray
band because of the high temperatures attained by the in-falling gas; however,
X-rays from space are absorbed by the Earth atmosphere. Since
its launch in 1999, the Chandra X-ray Observatory has contributed
substantially to our understanding of accreting black holes in the Milky Way
as well as external galaxies. This friendly site is a collage of Chandra
images, all equipped with accessible explanations. (E)

\item
{\url{gallery.spitzer.caltech.edu/Imagegallery/subcat.php?cat=Astronomical_Images&subcat=Galaxies_and_the_Universe}},\newline
{\bf Spitzer Space Telescope Images: Galaxies and the Universe}. The last of
NASA's Great Observatories to fly, Spitzer is a flying infrared telescope: UV/X-rays from
accreting supermassive black holes can be partly absorbed by surrounding dust
and re-emitted at lower frequencies. Here is how AGN would
look if human eyes were sensitive to the same frequencies as
microwaves. (E)
\end{enumerate}

\section{Statics,  Dynamics, and Thermodynamics of Black Holes}
\label{SDT}

This section addresses the well-established mathematical physics of black holes.  We have attempted to impose some structure on this highly interconnected body of knowledge by dividing the material as follows:  Section \ref{statics} deals with the statics of black holes, including the construction of stationary solutions, no-hair and uniqueness theorems, some work on hairy black holes, and definitions of horizons.  Perturbations, quasi-normal modes, and stability of stationary solutions is addressed in section \ref{pert}.  We have chosen to include the important subject of extreme-mass in-spirals and the calculation of gravity-wave production by post-Newtonian expansions in this section.  Section \ref{nonlin} then considers the strongly nonlinear dynamics of black holes associated with critical phenomena or with black-hole collisions.  Most of our discussion of numerical GR can be found here, as is the associated material on gravity waves.  Section \ref{thermo} contains resources on black-hole thermodynamics, quantum-field theory in curved spacetime, and the associated description of Hawking radiation.  More microscopic treatments of Hawking radiation and black-hole entropy are postponed to section \ref{micro}.  Sections \ref{statics}-\ref{thermo} focus primarily on black holes in $d=4$ spacetime dimensions; resources on black holes in other spacetime dimensions have been collected in section \ref{dnot4}.  We have chosen to include discussions of possible black-hole production at particle colliders in this latter section.

The reader should be aware of a few peculiarities that result from this organization.  Resources on gravity-wave production by black holes appear in both sections \ref{pert} and \ref{nonlin}.  Resources on isolated and dynamical horizons appear in sections \ref{statics}, \ref{nonlin}, \ref{thermo}, and also in section \ref{loop}.   There is also much overlap between section \ref{dnot4} (black holes in $d \neq 4$ dimensions) and sections \ref{thermo}, \ref{strings}, and \ref{last}, and between sections \ref{thermo} (thermodynamics) and \ref{other} (where we have chosen to locate discussions of entropy bounds and the proposed holographic principle, as well as Jacobson's work (ref. \ref{EEOS}) on the Einstein equation of state).  We apologize to the reader for any confusion this may cause.

\subsection{Statics}
\label{statics}

\begin{enumerate}[{\bf 1{.}}]

\addtocounter{enumi}{86}

\item { 'No-Hair' Theorems: Folklore, Conjectures, Results}, Chrusciel, P.T.,  in {Differential Geometry and Mathematical Physics}, J Beem, and K. L. Duggal., eds., vol 170, pp. 23-49 (Am. Math. Soc., Providence, 1994) [arXiv:gr-qc/9402032] .
{ ``Uniqueness of Stationary, Electro-Vacuum Black
Holes Revisited,''} P.T. Chrusciel, Helv. Phys. Acta, 69, 529-552 (1996). For a
related online version see:  { ``Uniqueness of
Stationary, Electro-Vacuum Black Holes Revisited,''} P.T. Chrusciel (October, 1996)
[arXiv:gr-qc/9610010  ].  An important property of
black holes is that, when left to themselves, they settle down to
solutions described by only a few parameters, most importantly their
mass $M$ and angular momentum $J$.  In other words, the surface of a
black hole {\it must} be very smooth; even in principle, the surface
of a black hole cannot support mountains, valleys, or other
complicated features.  One says that black holes ``have no hair.''
These papers review and complete classic work on the no-hair
theorems. (A)

\item {  ``Stationary Black Holes: Uniqueness and Beyond,''}
Markus Heusler, Living Reviews in Relativity (1998),
{\url{relativity.livingreviews.org/Articles/lrr-1998-6/index.html}}.
In sufficiently complicated settings, the number of parameters
needed to describe black holes increases and black holes can support
a limited amount of  ``hair."  This review describes the original
no-hair theorems as well as some of the more complicated hairy black
holes. See also ref. \ref{NA2} for more on non-abelian black holes. (A)

\item  \label{IB} { ``Black hole boundaries,''}
I.~Booth,
  Can.\ J.\ Phys.\  {\bf 83}, 1073-1099 (2005)
  [arXiv:gr-qc/0508107].
  Black holes are traditionally defined as regions of space-time from which in principle signals cannot propagate to infinity.  Such definitions clearly break down in cosmological settings, where appropriate asymptotic regions may not exist. Booth reviews recent work on alternative ``quasi-local'' definitions of  black-hole
  horizons. See also refs. ref{IH1},\ref{AK},\ref{Ann}. (A)

\end{enumerate}

\subsection{Perturbative Black-Hole Dynamics}
\label{pert}

 \begin{enumerate}[{\bf 1{.}}]
\addtocounter{enumi}{89}

\item \label{QN1} { ``Quasi-Normal Modes of Stars and Black Holes,''} K.
D. Kokkotas and B. Schmidt, Living Reviews in Relativity (1999),
{\url{http://relativity.livingreviews.org/Articles/lrr-1999-2/}}.
When a black hole is perturbed, it ``rings'' with a characteristic set
of frequencies.  However, the amplitude at each frequency decays as
energy falls through the horizon and radiates to infinity.  As a result, the solutions
associated with this ringing are called ``quasi-normal'' modes.
Kokkotas and Schmidt review such perturbations for both black holes
and relativistic stars, and in the process review the stability of
black holes.  See also ref. \ref{QN2}. (A)

\item \label{QN2} { ``Quasi-Normal Modes and Gravitational Wave Astronomy,''}  V.~Ferrari and L.~Gualtieri,  Gen.\ Rel.\ Grav.\  {\bf 40}, 945-970 (2008)
  [arXiv:0709.0657 [gr-qc]].
A review of quasi-normal modes with emphasis on possible detection
of the associated gravitational waves.   See also ref. \ref{QN1}. (A)

  \item  \label{EMRI1} { ``Extreme Mass Ratio Inspirals: LISA's unique probe of black hole gravity,''}
K.~Glampedakis, Class.\ Quant.\ Grav.\  {\bf 22}, S605-S659 (2005)
[arXiv:gr-qc/0509024]. The proposed LISA gravitational-radiation
detector is expected to be sensitive to gravity waves produced
when stars or stellar-mass black holes spiral into supermassive black holes.  In such cases the extreme ratio between the two black-hole masses
provides a small parameter that allows controlled calculations of
both these gravitational waves and the back-reaction on the small
body's orbit. This review begins with free-particle motion on the
black-hole background and builds up toward more complicated
calculations. (A)

\item \label{EMRI2} { Analytic Black Hole Perturbation Approach to Gravitational Radiation,''}
M. Sasaki and H. Tagoshi, Living Reviews in Relativity (2003),
{\url{relativity.livingreviews.org/Articles/lrr-2003-6/index.html}}. A
slightly more advanced review of extreme-mass in-spirals describing
both the relevant perturbative and post-Newtonian techniques. (A)

\item { ``Resource Letter GrW-1: Gravitational waves,''}
J.~M.~Centrella, Am.\ J.\ Phys.\ {\bf 71}, 520-525 (2003) [arXiv:gr-qc/0211084].
Centrella's Resource Letter is an excellent guide to the pre-2003 literature
on the production of gravity waves by events involving black holes. (E/I/A)

\item  {\bf Black Holes: The Membrane Paradigm},
by K. S. Thorne, D. A. MacDonald, and R. H. Price (Yale University
Press, 1986). This compilation of classic papers shows that black-hole horizons respond to outside perturbations like a
two-dimensional viscous fluid or membrane.  The horizon conducts
electricity but (at least classically) does not conduct heat.  As a
result, horizons are described by both the Navier-Stokes equations
and Ohm's law. (A)

\item { ``The internal structure of black holes},
W. Israel, in {Black holes and relativistic stars} edited by R.
Wald (U. of Chicago Press, Chicago, 1998) 137-151.  Analytic
solutions for charged and rotating black holes contain a so-called
Cauchy horizon at which predictability appears to break down.
However, this Cauchy horizon is unstable to the formation of a
singularity (where quantum effects clearly become relevant).  This
concise summary of black-hole interiors provides a good introduction
and guide to the pre-1998 literature on Cauchy-horizon instabilities
and the resulting phenomenon of mass inflation.  See also { ``The interior of charged black holes and the problem of uniqueness in general relativity,''} M. Dafermos,
Communications on Pure and Applied Mathematics,  {\bf 58}  445-504,  (2005) [gr-qc/0307013].
(A)

\item \label{NA2}  { Internal Structure of Black Holes and Spacetime Singularities:
Proceedings}, Annals of the Israel Physical Society vol. 13,  edited
by L. M. Burko and A. Ori (Institute of Physics Publishing, London,
1998). This volume contains a number of excellent articles on
subjects ranging from Cauchy-horizon instabilities (in diverse
contexts) to the internal structure of non-abelian black holes. (A)

\end{enumerate}

\subsection{Nonlinear Black-Hole Dynamics: Numerical GR and Black-Hole Collisions}
\label{nonlin}

 \begin{enumerate}[{\bf 1{.}}]
\addtocounter{enumi}{97}

\item { ``Critical Phenomena in Gravitational Collapse,''} C.
Gundlach, Living Reviews in Relativity (1999),
{\url{relativity.livingreviews.org/Articles/lrr-1999-4/index.html}}.
In 1993,  Choptuik discovered
a kind of critical phenomenon associated with black-hole formation in certain contexts.  For spherically symmetric scalar
fields, he considered generic 1-parameter families of initial data
that interpolate between weakly gravitating data (which does not
form a black hole) and strongly gravitating data (which does).  At
the threshold of black-hole formation one finds a complicated,
singular, scale-periodic solution while black holes of arbitrarily
small mass arise close to this threshold. Related phenomena were
later found in many systems.  Gundlach reviews the modern
understanding of this field. (A)

\item
{  ``Binary Black Hole Coalescence,''}   F.~Pretorius,
  arXiv:0710.1338 [gr-qc].
The coalescence of two black holes of comparable mass is a difficult
problem, best treated by numerical techniques.  (See refs. \ref{EMRI1},
\ref{EMRI2} for the case of extreme-mass ratios.)  After many years
of effort, such methods have recently become quite successful at
calculating the properties of the final black hole and producing
detailed waveforms for gravitational radiation. Pretorius reviews
these techniques, their results, and the implications for
astrophysics and other fields. (A)

\item { ``Initial Data for Numerical Relativity,''} Greg Cook, Living Reviews in Relativity,
lrr-2000-5,
relativity.livingreviews.org/Articles/lrr-2000-5/index.html
Extracting useful information from numerical simulations requires
starting with appropriate initial data.  Cook describes the current
methods and their limitations, with an eye toward future progress.
(A)

\item { ``Introduction to Isolated Horizons in Numerical
Relativity,''}
  O.~Dreyer, B.~Krishnan, D.~Shoemaker and E.~Schnetter,
  Phys.\ Rev.\  D {\bf 67}, 024018, 14pp. (2003)
  [arXiv:gr-qc/0206008].
Dreyer {\it et al} describe methods for using isolated-horizon techniques
(see refs. \ref{IB},\ref{IH1}, \ref{AK}, \ref{Ann})  to compute the mass $M$ and angular
momentum $J$ of black holes in numerical simulations. (A)

\end{enumerate}

\subsection{Black-Hole Thermodynamics and Hawking Radiation}

\label{thermo}

See also refs. \ref{2d},  \ref{TeV1}-\ref{TeV3}.

\begin{enumerate} [{\bf 1{.}}]

\addtocounter{enumi}{101}

\item \label{FirstLaw}
{ ``The Four Laws of Black Hole Mechanics,''} J. M. Bardeen, B.
Carter, and S. W. Hawking,  Commun. Math. Phys., {\bf 31}, 161-170,
(1973). This seminal paper derives a ``first law of black-hole
mechanics'' with striking similarities to the first law of
thermodynamics.  The authors argue that black holes also satisfy
analogues of the 0th, 2nd, and 3rd laws of thermodynamics.   The
idea that black holes are thermodynamic systems is now widely
regarded as a fundamental clue to the nature of quantum gravity. (A)

\item \label{entropy} { ``Black Holes and Entropy,''} J. D. Bekenstein, 
Phys. Rev. D, {\bf 7}, 2333-2346 (1973).  { ``Generalized
Second Law of Thermodynamics in Black-Hole Physics,''} J.D. Bekenstein,   Phys. Rev. D,
{\bf 9}, 3292-3300 (1974). These foundational works suggest that the area
of a black hole measures its entropy and formulate the generalized
second law of thermodynamics:  that the summed entropy of matter
plus black holes should not decrease. (A)

\item \label{Wald} { ``The thermodynamics of black holes,''}
R. Wald, Living Rev.\ Rel.\  {\bf 4}, 6 pp. 46 (2001)
[arXiv:gr-qc/9912119].
relativity.livingreviews.org/Articles/lrr-2001-6/index.html .
Wald provides a comprehensive survey of black-hole thermodynamics and the generalized second law.
A detailed guide to the literature is presented, to which the reader is referred for derivations.  Hawking radiation and associated issues are also discussed.  (I/A)

\item \label{IH1}
{``Black Hole Mechanics},'' A. Ashtekar, in {\bf Encyclopedia of
Mathematical Physics}, (Elsevier, New York, 2006) 300-305 .  Also available at
cgpg.gravity.psu.edu/people/Ashtekar/articles/\newline
bhm.pdf Ashtekar briefly reviews the standard formulation of black-hole mechanics and then describes the more local isolated- and
dynamical-horizon formulations.  See also refs. \ref{IB},\ref{AK},\ref{Ann}. (A)

\item \label{AK} { ``Isolated and Dynamical Horizons and Their Applications,''} A. Ashtekar and B. Krishnan, Living Reviews in Relativity, lrr-2004-10.
relativity.livingreviews.org/Articles/lrr-2004-10/  This more thorough review begins with the basics of the isolated and dynamical horizon quasi-local formalism for black-hole thermodynamics, but proceeds to address applications in numerical GR, mathematical physics, and quantum gravity.  (A)

\item \label{HawkRad} { ``Particle Creation by Black Holes,''} S. W. Hawking,  Commun. Math. Phys., {\bf 43}, 199-220
(1975). Hawking shows that, due to quantum effects, black holes emit
thermal radiation at (very low) temperatures set by their so-called
surface gravity. This is the original paper on such Hawking
radiation. The calculation identifies the black-hole entropy with
$A/4\ell_p^2$ where $A$ is the event-horizon area and $\ell_p$ is the so-called Planck length.  (A)

\item \label{TJRev}
{ ``Introduction to quantum fields in curved spacetime and the
Hawking effect,''}
 T.~Jacobson in {\bf Lectures on Quantum Gravity}, edited by A.
 Gomberoff and D. Marolf (Springer, New York, 2005).
  [arXiv:gr-qc/0308048.]
Based on lectures given at a summer school, Jacobson's notes provide
a very physical introduction to quantum fields in curved spacetime
and the Hawking effect.  This is an excellent reference for students
new to the subject. Cosmological particle creation, transplanckian
questions, and other issues are also addressed. The focus is on
Lorentzian techniques. (A)

\item \label{SRRev} { ``Black hole thermodynamics,''}
S.~F.~Ross,  arXiv:hep-th/0502195.
Also based on lectures at a summer school, Ross's notes are a good
complement to ref. \ref{TJRev} above.  This work includes a brief
introduction to black holes and a review of classical black-hole
thermodynamics.  Ross also discusses quantum-field theory in curved
spacetime, but emphasizes Euclidean techniques. (A)

\item \label{BD} {\bf Quantum Fields in curved space}, N. D. Birrell and P. C. W. Davies (Cambridge University Press, Cambridge, 1982).
A classic textbook on quantum-field theory in curved spacetime. This
text begins with the basics and works up through the derivation of
Hawking radiation. (A)

\item \label{WaldQFT} {\bf Quantum Field Theory in Curved Spacetime and Black Hole Thermodynamics}, R. M. Wald (U. Chicago Press, Chicago,
1994). This brief text quickly summarizes these fields as of 1994.
The approach is somewhat more mathematical than
refs. \ref{TJRev}-\ref{BD}. (A)

\item
{  ``Path Integral Derivation Of Black Hole Radiance,''}
J.~B.~Hartle and S.~W.~Hawking,
  Phys.\ Rev.\  D {\bf 13}, 2188-2203 (1976);
{ ``Action Integrals and Partition Functions in Quantum
Gravity,''} G. Gibbons, and S. W. Hawking, Phys. Rev. D, {\bf 15},
2752-2756 (1977). Together, these two papers introduce Euclidean
techniques into black-hole thermodynamics.  Hartle and Hawking use
analytic continuation to derive Hawking radiation, while Gibbons and
Hawking introduce the use of Euclidean solutions to approximate the
gravitational partition functions. (A)

\item \label{modify}  { ``On the origin of the particles in black hole evaporation,''}
 R.~Schutzhold and W.~G.~Unruh,
  arXiv:0804.1686 [gr-qc];  The above references derive Hawking radiation within the framework of locally Lorentz-invariant quantum-field theory on a fixed background spacetime.  This is certainly not the complete description of our universe, and some modification of Hawking radiation is expected.   The effect of breaking the local Lorentz symmetry has been explored in some detail, and Schutzhold and Unruh give the best understanding to date.  The present evidence (as does ref. \ref{TJRev} above) is that Hawking radiation is essentially a low-energy phenomenon, and should not receive significant corrections.   (A)

\item \label{Helfer}    { ``Do black holes radiate?,''} A.~D.~Helfer,
  Rept.\ Prog.\ Phys.\  {\bf 66}, 943-1008,  (2003)
  [arXiv:gr-qc/0304042].  As a counterpoint to ref. \ref{modify}, we include  Helfer's review of arguments that Hawking radiation might receive large corrections from quantum-gravity effects.  The reader should note, however, that certain arguments in this work are out of date in view of ref. \ref{TJRev} and ref. \ref{modify}. (A)

\item \label{higher} { ``Black hole entropy is the Noether charge,''} R. M. Wald, Phys. Rev. D {\bf 48}, 3427-3431 (1993) [arXiv:gr-qc/9307038]. { ``Some properties of Noether
charge and a proposal for dynamical black hole entropy,''}  V. Iyer
and R. M. Wald,  Phys. Rev. D {\bf 50}, 846-864 (1994) [arXiv:gr-qc/9403028].
{ "A Comparison of Noether charge and Euclidean methods for
computing the entropy of stationary black holes,} Phys. Rev. D {\bf 52},
4430-4439 (1995) [arXiv:gr-qc/9503052].   These works generalize the
first law of black-hole thermodynamics to theories more complicated
than just Einstein gravity plus matter.   Higher derivative theories are included.  (A)

\end{enumerate}

\subsection{Black Holes in Dimensions $d \neq 4$}

\label{dnot4}

\begin{enumerate}[{\bf 1{.}}]

\addtocounter{enumi}{115}

\item \label{BTZ} { ``The Black hole in three-dimensional space-time,''}
  M.~Banados, C.~Teitelboim and J.~Zanelli,
  Phys.\ Rev.\ Lett.\  {\bf 69}, 1849-1851 (1992)
  [arXiv:hep-th/9204099];
 {   ``Geometry of the (2+1) black hole,''}  M.~Banados, M.~Henneaux, C.~Teitelboim and J.~Zanelli,  Phys.\ Rev.\  D {\bf 48}, 1506-1525 (1993)
  [arXiv:gr-qc/9302012].
  In three dimensions, black holes arise only in the presence of a negative cosmological constant
  ($\Lambda < 0$). These so-called BTZ black holes are particularly
  simple, as they are quotients of 2+1 Anti-de Sitter space.
  The above works introduce such solutions and study their properties.
  (I/A)

\item \label{2d} { ``Dilaton gravity in two dimensions,''}
   D.~Grumiller, W.~Kummer and D.~V.~Vassilevich,
  Phys.\ Rept.\  {\bf 369}, 327-430 (2002)
  [arXiv:hep-th/0204253].
  Many calculations simplify in 1+1 dimensions.  While there is
  no meaningful theory of Einstein gravity in 1+1 dimensions, the
  so-called dilaton gravity theories capture some of the same
  physics and can be related to Einstein gravity in higher
  dimensions.  This work reviews such models, with emphasis on black-hole solutions, Hawking radiation, and quantum effects. (A)

\item \label{ER1}
  { ``Black Holes in Higher Dimensions,''} R.~Emparan and H.~S.~Reall,
  arXiv:0801.3471 [hep-th].  To appear in Living Reviews in Relativity.
  The space of black-hole solutions and behaviors becomes much
  richer in $d > 4$ dimensions where one finds black strings, black branes, and
  black rings in addition to the natural generalizations of 3+1
  black holes.  New behaviors arise as well, as some black objects
  are unstable and appear to lead to violations of certain cosmic
  censorship conjectures.  Emparan and Reall review these solutions, the
  techniques used to construct them, their thermodynamics, and
  what is known about their dynamics.  See also ref. \ref{instab}.  The section on open problems
  may be of particular interest for the motivated student. (A)

\item \label{ERRings}
{ ``Black rings,''} R.~Emparan and H.~S.~Reall,
  Class.\ Quant.\ Grav.\  {\bf 23}, R169-R197 (2006)
  [arXiv:hep-th/0608012].
This more-focused review addresses both the classical physics and
microscopic string theory of black rings. (A)

\item \label{instab}  { ``Instabilities of black strings and branes,"}
T.~Harmark, V.~Niarchos and N.~A.~Obers,
  Class.\ Quant.\ Grav.\  {\bf 24}, R1-R90 (2007)
  [arXiv:hep-th/0701022].
  While black holes in 3+1 dimensions are typically stable, black objects in higher dimensions exhibit interesting instabilities that may lead to violations of at least certain forms of the Cosmic Censorship Hypothesis.  This work reviews the classic Gregory-Laflamme instability of 4+1 black strings as well as many generalizations.  The correlated stability conjecture and its limitations are discussed. As in ref. \ref{ER1},  the section on open problems may be of particular interest for the motivated student. (A)

  \item \label{KolReview}
  { ``The phase transition between caged black holes and black strings: A
  review,''}
    B.~Kol,
  Phys.\ Rept.\  {\bf 422}, 119-165 (2006)
  [arXiv:hep-th/0411240].
In $d> 4$ dimensions, which sort of black-hole solution is most
stable can depend on the size of the box or ``cage'' in which it is
placed.  Large boxes favor black holes, while small boxes favor
black strings or branes. Kol reviews the complex structure
associated with the associated phase transitions. (A)

\item { ``Black holes and solitons in string theory,''} D.~Youm, Phys.\ Rept,\  {\bf 316}, 1-232 (1999) [arXiv:hep-th/9710046].
An encyclopedic review of black-hole and black-brane solutions in
string theory as of 1997. (A)

\item \label{stelle} { ``BPS branes in supergravity,''} K.~S.~Stelle, arXiv:hep-th/9803116, in {\bf 1997 Summer School in High Energy Physics and Cosmology}, edited by E. Gava {\it et al.} (World Scientific, New Jersey, 1998), pp. 29-127.
Stelle introduces brane solutions in supergravity, discussing super-symmetry, Kaluza-Klein reduction,
and low-velocity scattering of branes.  (A)

\item \label{marolf} { ``String/M-branes for relativists,''} D.~Marolf,
arXiv:gr-qc/9908045.  A brief introduction to stringy branes,
intended to convey some basic aspects of brane physics and
perspectives on string theory to those trained in GR.  (E/I/A)

\item \label{TeV1}
{ ``Black holes at future colliders and beyond: A review,''}
 G.~L.~Landsberg,
  arXiv:hep-ph/0211043.
  If our universe has more than four dimensions, then it may be
  possible to produce black holes at particle colliders,
  or they may be produced naturally in our atmosphere owing to the impacts of cosmic rays.
  Landsberg reviews the status of these ideas as of 2002, including
  the signatures of such events that could be used to detect them.
  (A)

\item \label{TeV2}
{ ``High-energy black hole production,''} S.~B.~Giddings,
  AIP Conf.\ Proc.\  {\bf 957}, 69-78 (2007)
  [arXiv:0709.1107 [hep-ph]].
A brief but more recent review of the possible production and
detection of TeV-scale black holes.  Includes a discussion of open
problems. (A)

\item \label{TeV3} {  ``Black holes in theories with large extra dimensions: A
review,''}
  P.~Kanti,
  Int.\ J.\ Mod.\ Phys.\  A {\bf 19}, 4899-4951 (2004)
  [arXiv:hep-ph/0402168].
A more thorough review of mini black holes in models with large
extra dimensions, focussing on their decay processes. (A)

\item { ``Braneworld black holes in cosmology and astrophysics,''
 } A.~S.~Majumdar and N.~Mukherjee,
  Int.\ J.\ Mod.\ Phys.\  D {\bf 14}, 1095-1129 (2005)
  [arXiv:astro-ph/0503473].
A review of larger black holes in models with large extra dimensions and their astrophysical implications. (A)

\end{enumerate}

\section{Black-Hole Microphysics}

\label{micro}

The resources below address the deep microphysics of black holes.  The reader should be aware that most of these works make certain assumptions about the quantum nature of gravity and spacetime that are beyond the reach of current experiments.  We have attempted to sort this material on the basis of these assumptions into string theory (section \ref{strings}), loop quantum gravity (section \ref{loop}), and other (section \ref{other}).   Section \ref{other} includes not only other approaches to quantum gravity, but also discussions of entropy bounds, the proposed holographic principle, black-hole complementarity, and Jacobson's work on the Einstein equation of state ref. \ref{EEOS}.

Certain overlaps and ambiguities are of course unavoidable.  The string-theory material is closely related to that of sections \ref{dnot4} (higher dimensions) and \ref{last} and has at least historical connections to some material in section \ref{other}, while the loop gravity work below is intimately tied to the treatments of isolated and dynamical horizons refs. \ref{IB},\ref{IH1}, and \ref{AK} (which we have chosen to list again as ref. \ref{AK2}).  There is also much overlap between sections \ref{other} and \ref{thermo}.

\subsection{Black-Hole Microphysics in String Theory}

\begin{enumerate}[{\bf
1{.}}]\addtocounter{enumi}{128}
\label{strings}

\item
{\bf String Theory, Vols. 1 \& 2}, J. Polchinski (Cambridge
University Press,Cambridge, 1998). {\bf D-branes}, C. Johnson
(Cambridge University Press, Cambridge, 2003). {\bf String Theory
and M-Theory: A Modern Introduction}, K. Becker, M. Becker, and J.
Schwarz (Cambridge University Press, Cambridge, 2007).  {\bf String theory in a Nutshell}, E. Kiritsis (Princeton University Press, Princeton, 2007).  These
standard string-theory textbooks each include at least a chapter on
black holes in string theory.  They are designed for readers with a
working knowledge of quantum-field theory and gauge theories in
addition to the usual undergraduate physics material. (A)

\item   { ``Resource letter: The nature and status of string
theory,''}
 D.~Marolf,
  Am.\ J.\ Phys.\  {\bf 72}, 730-741 (2004)
  [arXiv:hep-th/0311044].
This Resource Letter contains a more comprehensive guide to the
string theory literature, including the pre-2004 literature on
stringy black holes. (A)

\item { ``TASI lectures on black
holes in string theory,''} A.~W.~Peet, arXiv:hep-th/0008241, in {\bf
Strings,Branes, and Gravity: TASI 99: Boulder, Colorado, 31 May - 25
June 1999}, edited by J. Harvey, S. Kachru, and E. Silverstein
(World Scientific, Singapore, 2001), pp.  353-433. Peet's thorough
introduction to black holes and branes takes the reader up though a
stringy treatment of black-hole entropy and Hawking radiation. (A)

\item { ``The Quantum Physics Of Black Holes: Results From String Theory,''}
 S.~R.~Das and S.~D.~Mathur, Ann.\ Rev.\ Nucl.\ Part.\ Sci.\  {\bf 50}, 153-206 (2000) [arXiv:gr-qc/0105063].
This review takes the reader as directly as possible to the stringy
description of black-hole entropy and Hawking radiation.  (A)

\item { ``Microscopic formulation of black holes in string theory,''}
J.~R.~David, G.~Mandal, and S.~R.~Wadia, Phys.\ Rept,\  {\bf 369},
549-686 (2002) [arXiv:hep-th/0203048].  The authors review the
calculation of black-hole properties from string theory for the case
of the so-called nearly-extreme D1-D5 black hole.  Many important details of
the relevant gauge and conformal field theories are discussed,
leading to black-hole thermodynamics and Hawking radiation.  (A)

\item { ``Black holes in string theory,''} J.~M.~Maldacena,
arXiv:hep-th/9607235.   Maldacena's Ph.D. thesis was not written as
an introduction for outsiders, but does contain detailed treatments
of gauge-theory aspects relevant to the stringy counting of black-hole
entropy that are hard to find in other sources. Other useful aspects
of brane and black-hole physics are also described. (A)

\item
{ ``The fuzzball proposal for black holes: An elementary
review,''} S.~D.~Mathur,
  Fortsch.\ Phys.\  {\bf 53}, 793-837 (2005)
  [arXiv:hep-th/0502050].
Mathur reviews a particular set of speculations about black-hole
microphysics in string theory. This proposal is by no means
established, but has attracted significant interest. (A)

\end{enumerate}

\subsection{Black-Hole Microphysics in Loop Quantum Gravity}
\label{loop}

\begin{enumerate}[{\bf 1{.}}]

\addtocounter{enumi}{135}

\item \label{Ann} {  ``Interface of general relativity, quantum physics and statistical
  mechanics: Some recent developments,''} A.~Ashtekar,
  Annalen Phys.\  {\bf 9}, 178-198 (2000)
  [arXiv:gr-qc/9910101].
 Ashtekar provides  an introductory overview of both isolated and dynamical horizons and black-hole entropy in loop quantum gravity.  This work attempts to be as nontechnical as possible, and makes a good first introduction to the subject. (I/A)

\item \label{LBH} { ``Quantum geometry and black-hole entropy,''}
  A.~Ashtekar, J.~Baez, A.~Corichi and K.~Krasnov,
  Phys.\ Rev.\ Lett.\  {\bf 80}, 904-907 (1998)
[arXiv:gr-qc/9710007]. Microstates of black holes are counted in
loop quantum gravity.  Their number agrees with the
Bekenstein-Hawking entropy up to issues associated with the
so-called Immirzi parameter. (A)

\item { ``Quantum horizons and black-hole entropy: Inclusion of distortion and
  rotation,''}  A.~Ashtekar, J.~Engle and C.~Van Den Broeck,
  Class.\ Quant.\ Grav.\  {\bf 22}, L27-L34 (2005)
  [arXiv:gr-qc/0412003].
The state counting of ref. \ref{LBH} is generalized to include
distorted and rotating black holes. (A)

\item \label{AK2} { ``Isolated and Dynamical Horizons and Their Applications,''} A. Ashtekar and B. Krishnan, Living Reviews in Relativity, lrr-2004-10.
relativity.livingreviews.org/Articles/lrr-2004-10/  Ashtekar and Krishnan review the loop quantum gravity description of black holes and provide a thorough discussion of background material.  (A)

\end{enumerate}

\subsection{Other Microphysics and Black Hole Entropy}

\label{other}

\begin{enumerate}[{\bf 1{.}}]

\addtocounter{enumi}{139}

\item \label{carlip} { ``Black Hole Thermodynamics from Euclidean Horizon Constraints''
  },  S.~Carlip,
  Phys.\ Rev.\ Lett.\  {\bf 99}, 021301 4pp. (2007)
  [arXiv:gr-qc/0702107]. Carlip investigates whether the
  Bekenstein-Hawking entropy $A/4\ell_p^2$ might be determined by
  subtle symmetries of black holes, independent of the detailed
  microscopic theory of quantum gravity. (A)

\item {   ``Black Hole Entropy as Causal Links,''}
D.~Dou and R.~D.~Sorkin,
  Found.\ Phys.\  {\bf 33}, 279-296 (2003)
   [arXiv:gr-qc/0302009].
Within the causal-set approach to quantum gravity, Dou and Sorkin
show that the number of causal links crossing the horizon is
proportional to the black-hole area. (A)

\item \label{entangle} { ``The Entropy Of The Vacuum Outside A Horizon,''} R. D. Sorkin,  Gen. Rel. Grav.,
Proceedings of the GR10 Conference, Padova 1983, edited by B. Bertotti, F.
de Felice, A. Pascolini (Consiglio Nazionale della Ricerche, Roma,
1983) Vol. 2; { ``A Quantum Source Of Entropy For Black Holes,''}
L. Bombelli, R. K. Koul, J. H. Lee and R. D. Sorkin, Phys. Rev. D
{\bf 34,} 373-383 (1986);  { ``Entropy and area,''} M. Srednicki,  Phys. Rev.
Lett. {\bf 71,} 666-669 (1993) [arXiv:hep-th/9303048].  These works introduce
the idea that black-hole entropy might fundamentally describe
correlations in the vacuum across the horizon of the black hole. The
corresponding entropy diverges in quantum field theory but can be
made to agree with the Bekenstein-Hawking entropy $A/4 \ell_p^2$ by
imposing a cut-off at the Planck length. (A)

\item \label{atmosphere}  { ``On The Quantum Structure Of A Black
Hole,''}
G. 't Hooft, Nucl. Phys. B {\bf 256}, 727-745 (1985).  A calculation much like that of ref. \ref{entangle} is presented, but from a different perspective.  In this so-called brick-wall model, the entropy of a black hole is associated with states of quantum fields outside of, but close to the horizon. (A)

\item \label{Centangle}   { ``Black hole horizon fluctuations,''}
 A.~Casher, F.~Englert, N.~Itzhaki, S.~Massar and R.~Parentani,
  Nucl.\ Phys.\  B {\bf 484}, 419-434 (1997)
  [arXiv:hep-th/9606106].
{  ``How wrinkled is the surface of a black hole?,''} R. D. Sorkin, in {\bf Proceedings of the First Australasian Conference on General Relativity and Gravitation}, edited by D. Wiltshire, (University of Adelaide, Adelaide, 1996)  pp. 163-174.
arXiv:gr-qc/9701056.  { ``On the quantum width of a black hole
horizon,'' } D.~Marolf,
  Springer Proc.\ Phys.\  {\bf 98}, 99-112 (2005)
  [arXiv:hep-th/0312059].
We include these works as a counterpoint to \# \ref{entangle}, \#\ref{atmosphere}. They argue that a natural cut-off occurs at a larger length scale, so that the vacuum entanglement entropy is negligible compared with $A/4\ell_p^2$. (A)

\item \label{BS} { ``TASI lectures on the
holographic principle,''} D.~Bigatti and L.~Susskind,
arXiv:hep-th/0002044, in {\bf Strings,Branes, and Gravity: TASI 99:
Boulder, Colorado, 31 May - 25 June 1999}, edited by J. Harvey, S.
Kachru, and E. Silverstein (World Scientific, Singapore, 2001), pp.
883-933.   This surprisingly nontechnical work reviews a number of interesting but controversial ideas about black-hole microphysics ranging from black-hole complementarity to entropy bounds and the holographic principle: the assertion that the number of states of a region of space is determined by the area of its boundary.  The more established anti-de Sitter/conformal field theory correspondence is discussed as an example of these ideas.  (I)

\item \label{RB} { ``The
holographic principle,''} R.~Bousso,  Rev.\ Mod.\ Phys.\ {\bf 74},
825-874 (2002) [arXiv:hep-th/0203101]. Bousso provides a thorough
review of entropy bounds, and related issues. Applications and examples
are presented and some relevant black-hole physics is reviewed.   Much of this discussion is accessible to nonexperts.  (I)
\newline

\item \label{CHolog} { ``Acceleration Radiation And Generalized Second Law Of
Thermodynamics,''} W.~G.~Unruh and R.~M.~Wald,
  Phys.\ Rev.\  D {\bf 25}, 942-958 (1982).
 { ``Entropy Bounds, Acceleration Radiation, And The Generalized Second
 Law,''}
  W.~G.~Unruh and R.~M.~Wald,
  Phys.\ Rev.\  D {\bf 27}, 2271-2276 (1983).
{ ``On the status of highly entropic objects,''}
   D.~Marolf and R.~D.~Sorkin,
  Phys.\ Rev.\  D {\bf 69}, 024014 5pp. (2004)
  [arXiv:hep-th/0309218].
    { ``Notes on spacetime thermodynamics and the observer-dependence of
  entropy,''}  D.~Marolf, D.~Minic and S.~F.~Ross,
  Phys.\ Rev.\  D {\bf 69}, 064006 (2004)
  [arXiv:hep-th/0310022].
We include these works as a counterpoint to some of the arguments in refs. \ref{BS} and \ref{RB}.
See also ref. \ref{Wald}. (A)

\item { "Ultimate physical limits to computation,"} S. Lloyd, Nature {\bf 406} 1047-1054 (2000)  [arXiv:quant-ph/9908043]; { ``Black hole entropy and quantum information,''}
 M.~J.~Duff and S.~Ferrara,
Ê arXiv:hep-th/0612036.
Ê 
These works explore connections between black-hole physics and quantum information theory. (A)

\item \label{EEOS} { ``Thermodynamics of space-time: The Einstein equation of
state,''} T.~Jacobson,
  Phys.\ Rev.\ Lett.\  {\bf 75}, 1260-1263 (1995)
  [arXiv:gr-qc/9504004].
All horizons in general relativity share the basic features of black-hole thermodynamics.  This intriguing paper shows that the usual
derivations can be reversed: by assuming that the first law holds
for all horizons in its standard form, one can derive the full
dynamics of Einstein gravity. (A)

\end{enumerate}

\section{Connections to Nuclear and Condensed Matter Physics}

\label{last}

This brief section addresses connections between black-hole physics and that of the at first sight unrelated fields of nuclear and condensed-matter physics.    The resources below are of two types.  First, ref. \ref{analogue} reviews the use of established hydrodynamics and condensed matter physics to design analogues of black holes in certain fluid systems and Bose-Einstein condensates.  In contrast, the remaining works attempt to use black holes and a bit of string theory as mathematical tools to understand quark-gluon plasmas (ref. \ref{RHIC}) and condensed-matter physics (ref. \ref{CM}).  It will be very interesting to see how these ideas develop in the near future.

\begin{enumerate}[{\bf 1{.}}]

\addtocounter{enumi}{148}

\item \label{analogue} { ``Analogue gravity,''}
 C.~Barcelo, S.~Liberati and M.~Visser,
  Living Rev.\ Rel.\  {\bf 8}, 12, 151pp. (2005) 
  [arXiv:gr-qc/0505065].
  The authors review the vast literature on analogues of black holes in fluid and condensed matter systems.  Both classical properties of black holes and Hawking radiation are discussed. There are interesting connections to ref. \ref{modify}. (A)

\item  \label{RHIC} { ``String Theory and Quantum Chromodynamics,''}
  D.~Mateos,
  Class.\ Quant.\ Grav.\  {\bf 24}, S713-S739 (2007)
  [arXiv:0709.1523 [hep-th]].
The so-called gauge/gravity dualities of string theory imply that certain
theories of gravity can be used to calculate properties of what might appear
to be completely unconnected quantum-field theories.  The best-studied example
of this is the anti-de Sitter/conformal field theory correspondence (AdS/CFT).
Finite temperature effects in quantum-field theory are associated with
black holes in the gravitational theory.  Mateos reviews attempts over the
last few years to use this correspondence to understand the physics of
quark-gluon plasmas such as those currently being generated at the
Relativistic Heavy Ion Collider (RHIC) in Brookhaven, New York. Some background in gauge theory and string theory is required. (A)

\item \label{CM} { ``Quantum critical transport, duality, and M-theory,''}
  C.~P.~Herzog, P.~Kovtun, S.~Sachdev and D.~T.~Son,
  Phys.\ Rev.\  D {\bf 75}, 085020, 21pp. (2007)
  [arXiv:hep-th/0701036];
{ ``Theory of the Nernst effect near quantum phase transitions in condensed
  matter, and in dyonic black holes,''}
 S.~A.~Hartnoll, P.~K.~Kovtun, M.~Muller and S.~Sachdev,
  Phys.\ Rev.\  B {\bf 76}, 144502, 17pp. (2007)
  [arXiv:0706.3215 [cond-mat.str-el]].
  Strongly coupled quantum effects are believed to play an important role in certain condensed-matter systems, such as high-temperature superconductors.  Near a phase transition these effects should be described by a strongly coupled conformal field theory.  As such, they may be amenable to study via the AdS/CFT correspondence.  These papers begin what will surely be a long process of attempting to do so.  (A)

\end{enumerate}

$~$\\

\acknowledgements  
The authors thank Amitabh Virmani, Steve
Giddings and Marta Volonteri for help in locating certain references.  We also thank Jessica Wirts for assistance in tracking down certain bibliographic information.
E.G. is funded by NASA through a 
Hubble Fellowship grant from the Space Telescope Science Institute, which is 
operated by the Association of Universities for Research in Astronomy, 
Incorporated, under NASA contract NAS5-26555. D.M. was supported  in part by the US National Science
Foundation under Grant No.~PHY05-55669, and by funds from the
University of California.

\end{document}